# Topologically protected Dirac plasmons in a graphene superlattice

Deng Pan,[1,2] Rui Yu,[1] Hongxing Xu,[1,*] and F. Javier García de Abajo[2,3,*]

[1]*School of Physics and Technology, Wuhan University, Wuhan 430072, China*
[2]*ICFO-Institut de Ciencies Fotoniques, The Barcelona Institute of Science and Technology, 08860 Castelldefels (Barcelona), Spain*
[3]*ICREA- Institució Catalana de Recerca i Estudis Avançats, Passeig Lluís Companys 23, 08010 Barcelona, Spain*
∗Corresponding authors: hxxu@whu.edu.cn, javier.garciadeabajo@icfo.es

## Abstract

**Topological optical states exhibit unique immunity to defects and the ability to propagate without losses rendering them ideal for photonic applications. A powerful class of such states is based on time-reversal symmetry breaking of the optical response. However, existing proposals either involve sophisticated and bulky structural designs or can only operate in the microwave regime. Here, we propose and provide a theoretical proof-of-principle demonstration for highly confined topologically protected optical states to be realized at infrared frequencies in a simple 2D material structure—a periodically patterned graphene monolayer—subject to a magnetic field of only 2 tesla. In our graphene honeycomb superlattice structures plasmons exhibit substantial nonreciprocal behavior at the superlattice junctions, leading to the emergence of topologically protected edge states and localized bulk modes enabled by the strong magneto-optical response of this material, which leads to time-reversal-symmetry breaking already at moderate static magnetic fields. The proposed**

**approach is simple and robust for realizing topologically nontrivial 2D optical states, not only in graphene, but also in other 2D atomic layers, and could pave the way for realizing fast, nanoscale, defect-immune devices for integrated photonics applications.**

Topologically protected photonic states [1,2] exhibit remarkable robustness against disorder and backscattering, which make them extremely useful for the realization of defect-immune photonics devices. So far, such states have been explored based on reciprocal metamaterials [3,4] and photonic crystals [5,6]. But their realization has relied on sophisticated structural designs, which greatly complicate their fabrication and limit their miniaturization for nanoscale optical integration. Additionally, those topologically protected edge modes could couple to backscattering channels along opposite directions compatible with time-reversal symmetry (T-symmetry).

A more robust method to realize topologically protected photonic states consists in introducing T-symmetry breaking in periodic structures [7-11], which leads to the opening of a topologically nontrivial bandgaps at Dirac and degeneracy points. T-symmetry breaking can be introduced via dynamical modulation of the refractive index [10], although this approach is extremely challenging from the experimental viewpoint. A simpler alternative consists in using the magnetization of the material to induce T-symmetry breaking by exposing it to a static magnetic field. While this approach has been demonstrated in the microwave regime [9], the weak magneto-optical (MO) response of most materials at visible and infrared frequencies renders it difficult to achieve substantial T-symmetry breaking in these technologically important spectral ranges. Thus, the question remains how topologically protected photonic states can be robustly achieved in practice at high frequencies and their benefits exploited for the design of defect-immune, fast-speed nanophotonics devices.

Recent studies have shown that Dirac fermion (DF) systems possess a giant MO response in the infrared regime, such as the conducting surface of the topological insulators [12] and a monolayer graphene [13-15]. Moreover, the plasmons supported by DF systems [16-23] show deep-subwavelength confinement. This confinement makes the Dirac plamons extremely sensitive to external modulations [24], and possibly more susceptible to the MO response than using light plane waves incident on structureless graphene, therefore providing substantial T-symmetry breaking in the infrared regime. These properties render DF systems as an ideal platform to realize ultra-confined topologically protected plasmon modes at high frequencies. Although edge magnetoplasmons already show obvious nonreciprocal behavior [13-15], and at low frequencies the magnetoplasmons even form topologically protected states [25], more practical and versatile applications are opened by constructing topologically nontrivial bandgaps in the high frequency regime, and additionally steering plasmon propagation by using periodically patterned graphene.

Here we theoretically demonstrate that topologically protected plasmonic states can be robustly realized in DF superlattices constructed from single-layer graphene, based on the giant MO response of this material under exposure to static magnetic fields of only a few tesla. The superlattice is a honeycomb network constructed by graphene nanoribbons (Fig. 1a). The applied magnetic field induces asymmetry in the guided ribbon plasmon modes, thus resulting in directional coupling at the junctions of the structure. We show that, as a direct consequence of this directional coupling, localized modes are formed inside the superlattice, as well as topologically protected edge states at the boundary.

**Results**

**Structure and working principle**. We focus on graphene [26] because it sustains ultra-confined infrared plasmons that have been already observed in experiments [21,22]. Topological protection of plasmons in our proposed graphene superlattice can be intuitively understood as schematically illustrated in Fig. 1a. In contrast to previous designs based on photonic crystals, our structure consists of a network of waveguides (ribbons) in which topological protection emerges by analogy to the classical phenomenological picture of the quantum Hall effect. In this picture, free electrons in a Fermi gas follow circular cyclotron orbits under the influence of a magnetic field, and consequently they form a bulk insulating state and a topologically protected conducting state at the edge boundary. Although photons are charge-free, and consequently, a magnetic field cannot change their direction of motion, the splitting ratio of the plasmons at the junctions of the superlattice can be controlled by the MO effect due to the breaking of T-symmetry, leading to nonreciprocal directional coupling (Fig. 1b-d).

In the limit of total directional coupling (e.g., nearly 100% of the power exiting through the right side at each junction, as shown in the right panel of Fig. 1d), the guided waves circulate around single lattice hexagons and become localized vortex states in the interior of the superlattice, mimicking the cyclotron motion of free electrons (routes 1 and 2 in Fig. 1a). Additionally, a unidirectional and topologically protected guided mode is produced at the edge boundary (route 3 in Fig. 1a). In a general situation characterized by nonzero backward and leftward scattering at each junction (e.g., middle panel in Fig. 1d), localized bulk modes and a topologically protected edge wave are formed at a frequency within the bandgap opened by the T-symmetry breaking under magnetic field exposure [7] at the Dirac point of the honeycomb superlattice –these localized bulk modes and the topologically protected edge state exhibit analogous behaviors to the phenomenological picture in Fig. 1a, and further elaborated in our calculations below. We note that other periodic patterns different from the honeycomb structure, such as triangular or square lattices with degenerate band points, are equally applicable.

The above design for realizing topologically protected plasmonic states is universal for waveguide networks incorporating a MO material, but the graphene nanoribbons are ideally suited because radiative losses can be neglected due to the large mismatch between plasmon and photon wavelenghs. Additionally, experiments have demonstrated rather low propagation losses in this material [27,28]. More importantly, graphene exhibits a giant MO response in the infrared regime, so the supported plasmons are highly susceptible to the external magnetic field and allow us to realize the required directional coupling (Fig. 1b-d).

We use the finite-element method to numerically solve Maxwell's equations and calculate the plasmon dispersion relations and associated field distributions (see Methods). The response of graphene under a static magnetic field $\boldsymbol{B}$ is described by the Drude conductivity [13] as

$$\sigma_{xx}(\omega, B) = \frac{e^2 E_F}{\pi \hbar^2} \frac{\gamma - i\omega}{\omega_c^2 - (\omega + i\gamma)^2} \qquad \sigma_{xy}(\omega, B) = -\frac{e^2 E_F}{\pi \hbar^2} \frac{\omega_c}{\omega_c^2 - (\omega + i\gamma)^2} \qquad (1)$$

where $E_F$ is the Fermi energy, $\gamma$ is the plasmon damping rate, $\omega_c = e \, \boldsymbol{B} \cdot z \, v_F^2 / E_F$ is the cyclotron frequency [29] and $v_F \approx 10^6$ m/s is the graphene Fermi velocity. For simplicity, we ignore losses ($\gamma = 0$), which is reasonable in view of the long plasmon lifetimes (> 500 fs) observed in graphene [27,28]. The local dielectric formalism is a reasonable approximation because the nanoribbons are hundreds of nanometers wide, and consequently, the Fermi energy exceeds the plasmon energy in all our calculations, thus eliminating any dependence on the crystallographic orientation of atomic edges [30].

The two lowest-order plasmonic modes of the nanoribbon are bonding and antibonding combinations of induced charge pileup at the ribbon edges (Fig. 1b, upper insects) [31], with the second order mode showing a wavelength cutoff. Under an externally applied static magnetic field (Fig. 1b, dashed curves), the field distributions of the two modes become asymmetric, with field piling up toward different sides of the ribbon (Fig. 1b, lower insects). This nonreciprocal behavior is caused by the MO response, which is captured by the off-diagonal elements of the conductivity (Equation 1), similar to what happens for edge magneto-plasmons in a two-dimensional electron gas [32, 33] and graphene [34] (see Sec. I in Supplementary information (SI)).

Below the cutoff frequency of the second mode, the ribbon only holds a single mode, so symmetry breaking can lead to directional coupling at the ribbon junctions, which is required for the realization of topological protection in the superlattice (Fig. 1c-d). The unit cell of the superlattice is formed by two nanoribbon junctions with lengths of branches equal to $L/2$. For each junction, we calculate the scattering coefficients $S_{1,1\text{-}3}$, starting from a plasmon wave of unit amplitude launched on the terminal of branch 1 and going to ports at the branches 1-3 (Fig. 1c). The results confirm that without magnetic field the powers at the outputs 2 and 3 are equal due to geometrical symmetry, while an applied magnetic field of 5 T leads to clear directional coupling (Fig. 1d, for a photon energy of 0.06 eV) and the right-coupling efficiency exceeds 96% for a magnetic field of 15 T.

**Band diagrams of the graphene supperlattice.** The magnetic-field-induced directional coupling at the junctions leads to the opening of a Dirac point (Fig. 2). Separate bands at this point have nonzero Chern numbers [7], which yield topologically protected edge modes within the bandgap, with the number of edge modes equal to the Chern numbers of the bands below the gap. To obtain the band diagram of the superlattice, we first calculate the scattering coefficients of the junction structure illustrated in Fig. 1c for different frequencies and magnetic field strengths via FEM simulations (Fig. 2a). Energy conservation leads to the relation $|S_{11}|^2+|S_{12}|^2+|S_{12}|^2\approx1$ between the scattering coefficients, which is satisfied by our numerical simulations.

Using Bloch's theorem, we calculate the band diagram of the superlattice (Sec. II of the SI) – Fig. 2b shows the results for an excursion along symmetry points within the first Brillouin zone. Our results confirm that, when a magnetic field of 8 T is applied, a large bandgap opens at the Dirac points K and K' (Fig. 2b,d and f), which results in the formation of localized bulk modes within the band gap region – in contrast, no bandgap is present in the absence of magnetic field (Fig. 2b,c and e). The bandgap opened by the magnetic field is topologically nontrivial, as demonstrated by the nonzero Chern number (C=1) of the first band below the gap (see Sec. II of the SI for details on the calculation). This band has a nonzero Chern number by exchanging topological charge with upper bands separated by the Dirac point. The lower bandgap (in between orange and blue bands in Fig. 2b) is topologically trivial, corresponding to a zero sum of Chern numbers over all bands below 44 meV; the trivial character of this bandgap is evidenced by the fact that it does not close even when increasing the magnetic field up to 8 T. According to the bulk-edge correspondence principle [35], the Chern number C=1 of the first band below $E_0$ implies a unidirectional anticlockwise edge mode on the external boundary of the superlattice in the opened bandgap. The dispersion relations of the edge modes are revealed by the projected band diagrams, which are calculated along the zigzag (Fig. 2c,d) and armchair (Fig. 2e,f) directions for a superlattice of finite period ($N$=20) (see Sec. II of the SI). The dispersion curves (red curve in Fig. 2d) in the bandgap correspond to edge modes propagating on the upper and lower boundaries of the superlattice – these have unidirectional group velocities, which imply that these edge modes are topologically protected. The directions of the group velocities also indicate that the edge mode on the external boundary of a finite surperlattice is anticlockwise, in agreement with that predicted from the Chern number.

**Topologically protected localized and edge modes.** To reveal the localized mode and prove the topological protection of the edge modes in the proposed graphene superlattice, we construct a numerical network with honeycomb topology and simulate the field evolution using the scattering coefficients represented in Fig. 2a (see Sec. III in SI). The field distributions for a directional point excitation (indicated by the red arrows) at the center of the network with magnetic field of 8 T is shown in Fig. 3a. For a photon energy of $E_0$=56.37 meV, which lies in the band gap opened by the magnetic field, the directional coupling at the junctions causes the excitation power to circulate around a single lattice hexagon with enhanced intensity – this conforms the localization of the mode and corroborates the phenomenological picture we introduced in Fig. 1a. As this picture suggested, the localized vortex mode is the foundation for the topologically protected states, which can be clearly demonstrated by Fig. 3b. We now introduce a vacant defect by removing several nanoribbons from the center of the superlattice. Then, a directional point excitation adjacent to the defect generates a localized wave circulating around the defect unidirectionally without backscattering, which reveals the topological protection of the edge mode on the interior boundary of the superlattice.

To demonstrate that the edge mode on exterior boundary of the superlattice is topologically protected, we further cut the honeycomb network into a complex shape (Fig. 4a) and simulate the field evolution in successive time steps. In the presence of a $B$=8 T field, a point source of energy $E_0$ placed at the boundary of the network can generate the sought-after one-way boundary mode, which efficiently propagate energy over sharp corners without back reflection (Fig. 4b). When the magnetic field is reversed in sign, the direction of the one-way edge state is also reversed (Fig. 4c). Finally, when the magnetic field decreases to 2 T, the edge mode is still preserved, although it is less localized at the boundary (Fig. 4d).

**The effect of inelastic optical losses.** We further show that inelastic losses do not deteriorate topological protection of the plasmons in the structure of Fig. 4, and just induce an attenuation of propagation, as shown in Fig. 5a. More precisely, we introduce losses into the Drude model through a damping rate $\gamma=\mu E_F/ev_F^2$, where $\mu$ is mobility of graphene, for which we adopt the experimentally measured value in high-quality suspended graphene [36], which can be maintained by decorating it with h-BN. Figure 5a clearly shows that the excitation point source still generates unidirectional edge modes with good contrast between the two opposite directions, and that the edge modes retain a good capacity of topological protection. Inelastic losses only result in the attenuation of the modes along their propagation. In Fig. 5a, we show that edge plasmons can propagate over ~30 periods. Since the individual branch structures under consideration can serve as classical and quantum interferometers [37, 38, 39] or as logic gates [40, 41], the proposed graphene superlattice provides a versatile platform to explore various complex nonreciprocal optical computing functions.

Importantly, similar graphene superlattices of different sizes are equivalent according to basic electrostatic scaling laws of plasmons in 2D materials [42, 43] (see details in Sec. IV of the SI), which reveal the possibility to further miniaturize the graphene superlattice in order to increase the propagation distance in units of plasmon wavelengths. In particular, when the geometry of the graphene superlattice in Fig 4d shrinks by a factor of 2 and the magnetic field increases by a factor of $\sqrt{2}$, for a photon energy of $\sqrt{2}\,E_0$, the response of the superlattice and the field distribution is equivalent to that eFig. 4d. However, when considering inelastic losses (see Fig. 5b), the propagating distance increases considerably in units of the plasmon wavelength [31] compared with Fig. 5a, which means that the plasmon wave can propagate over many more periods in the network. Therefore, the noted scaling leads to a dramatic reduction in the effect of inelastic losses that should facilitate the design of practical devides.

**Discussion**

The realization of the topologically protected plasmonic states in the proposed graphene superlattice is experimentally feasible using state-of-the-art fabrication and measurement techniques, similar to those recently used to observe edge plasmon modes on graphene nanoribbons [44] (i.e., the elementary unit based on which we obtain topologically protected plasmons). To reduce inelastic losses, the sample can be fabricated by dielectric structuring rather than direct lithographic patterning of the graphene, thus relying on extended graphene encapsulated in hBN, for which the measured mobility exceeds the values here assumed [27]. Additionally, our results can be readily extrapolated to include the effect of the substrate, the size of the structure and the level of graphene doping via an electrostatic scaling law formulated for 2D structures [42,43] (see details in Sec. IV of the SI). An extension of this law to include the MO response reveals that the magnetic field strength required to achieve topological protection is below 1 tesla – such field strengths can be achieved using commercially available permanent magnets.

In conclusion, we have proposed and theoretically demonstrated how topologically protected plasmon modes can be realized robustly in single-layer graphene honeycomb surperlattice structures using experimentally attainable static magnetic fields. The key ingredients for this realization are the strong MO response of graphene, which induces T-symmetry breaking, and the ensuing directional coupling at geometrically symmetric junctions between nanoribbons in the lattice. This nonreciprocal propagation leads to the formation of localized bulk modes and a topologically protected edge mode. Although we have focused on graphene, we expect other kinds of 2D materials, such as topological insulators, semiconductor junctions and conducting 2D materials, to exhibit similar phenomena, for example by exploiting their long-lived phonon-polaritons. Our design provides an a simple, yet robust platform for fast speed, ultra-compact, nonreciprocal optical computing networks, thus paving the way toward realistic applications of topological photonics.

## Methods

**Numerical simulations.** We adopt a finite-element method (COMSOL) to numerically calculate the electromagnetic field distribution and plasmon dispersion relations of uniform graphene nanoribbons, as well as the reflection and left/right-transmission coefficients of plasmons at the nanoribbon junction structures. We model graphene as a thin layer of thickness $d$ and permitivity $e(w) = 1 + i s(w)/(e_0 d w)$, where $s(w)$ is the frequency-dependent 2D conductivity in the Drude model (equation (1)). In the simulation, we take $d$=0.5 nm as a reasonable value close to the $d \rightarrow 0$ limit. The scattering coefficients are calculated as the ratios between the complex amplitudes of the absorbed and input fields at ports on each of the three branches of the junction, as shown in Fig. 1c.

## Author contributions

D.P. and H.X. conceived the nonreciprocal propagation in nanoribbon junctions. F.J.G.d.A. proposed the topological protection in graphene superlattice. F.J.G.d.A. and D.P. worked out the theory. D.P. performed the simulation and wrote the paper. R.Y. collaborated in the calculation of the projected diagram and discussed the results. H.X. and F.J.G.d.A. supervised the projected and revised the paper.

## Additional information

Supplementary information is available in the online version of the paper. Reprints and permissions information is available online at www.nature.com/reprints.

Correspondence and requests for materials should be addressed to H.X. or F.J.G.d.A.

## Competing financial interests

The authors declare no competing financial interests.

# Figures

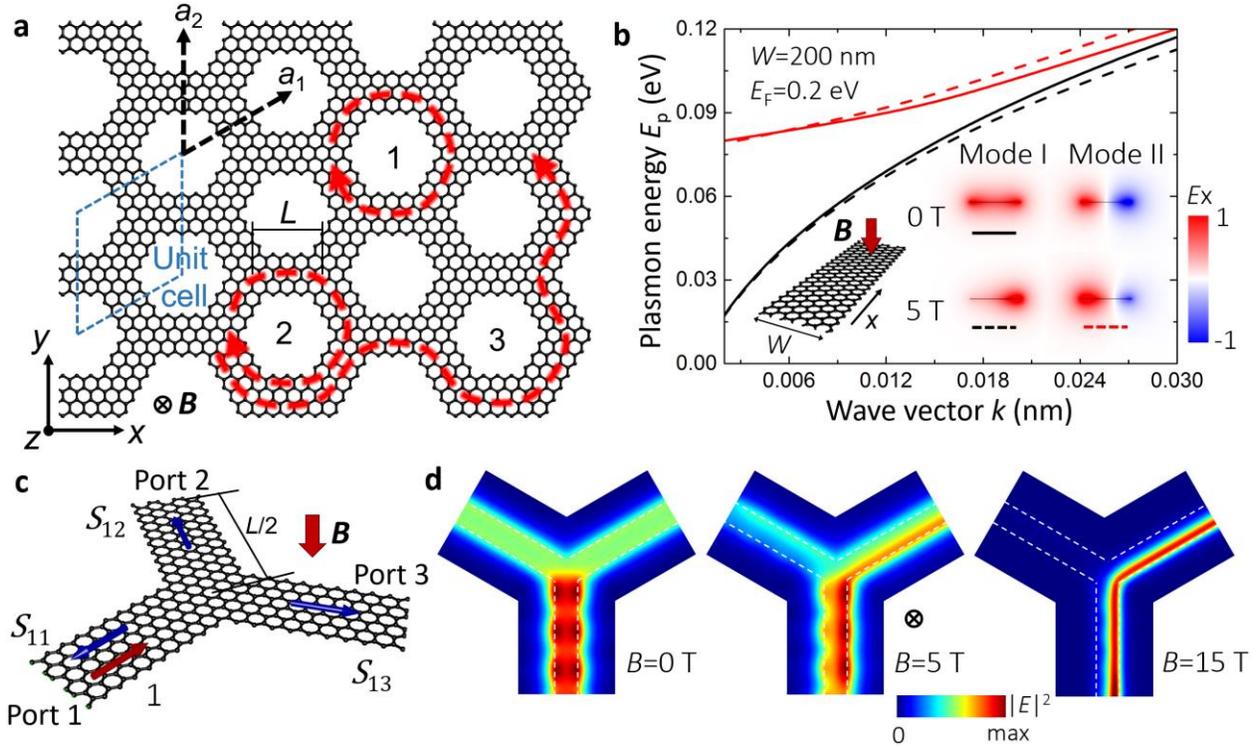

**Figure 1 | Topologically protected plasmons in a graphene superlattice resulting from directional coupling at the lattice junctions.** (**a**) Illustration of a honeycomb graphene superlattice with 100% right-coupling efficiency at each junction. Plasmons form localized vortex modes (1 and 2) and a unidirectional topologically protected edge state (3). (**b**) Dispersion relations (curves) and electric-field distributions (right insects, $E_p$=0.1 eV) of the two lowest-order modes sustained by a graphene nanoribbon (width $W$=200 nm, Fermi energy $E_F$=0.2 eV) with and without a normal static magnetic field $\boldsymbol{B}$. (**c**) Schematic of a graphene nanoribbon junction with plasmons incident from branch 1 (unit incident amplitude), and

scattered toward the three branches 1-3 with coefficients $S_{1,2\text{-}3}$. (**d**) Electric-field distributions (50 nm above graphene for $E_p$=0.06 eV) produced by plasmon scattering at the nanoribbon junction with and without magnetic field, assuming the same ribbon parameters as in **b** and neglecting inelastic losses. Dashed lines delineate the graphene edges.

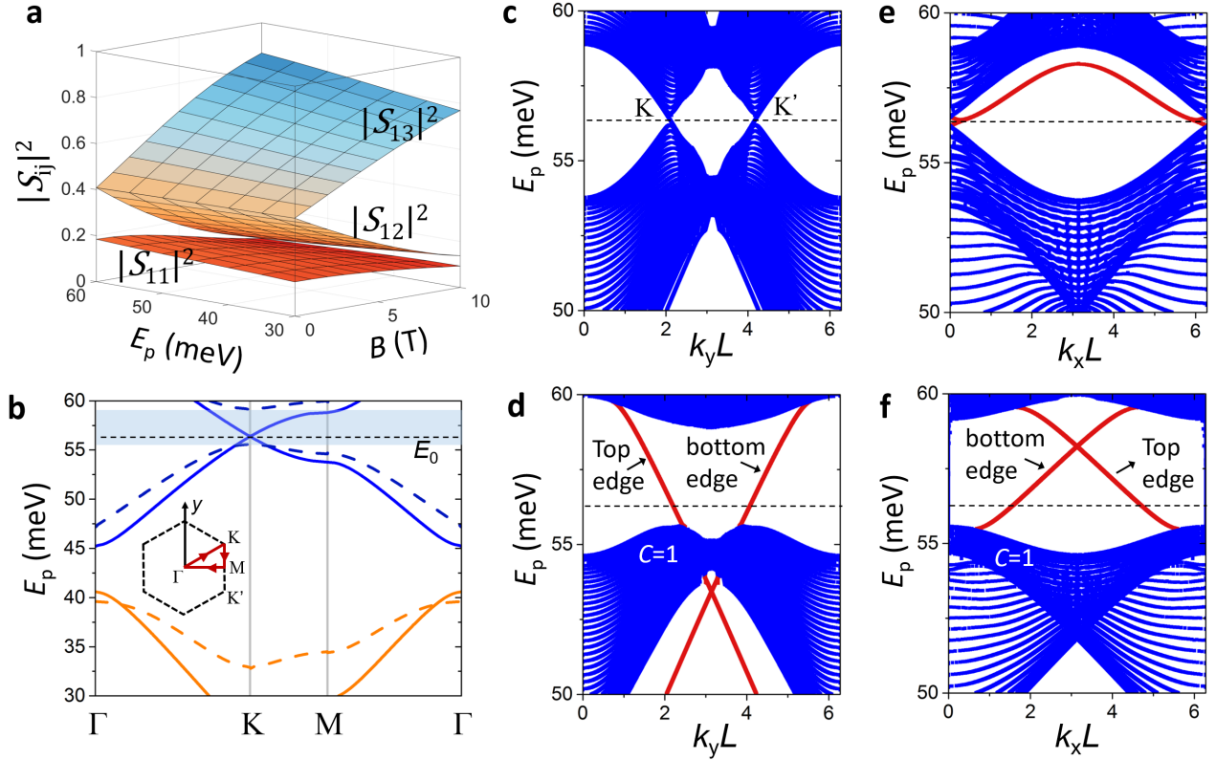

**Figure 2 | Plasmon band diagrams of the graphene superlattice.** (**a**) Squared amplitude of the junction scattering coefficients (see Fig. 1c) as a function of plasmon energy and magnetic-field strength $B$ for the same ribbon parameters as in Fig. 1d. (**b**) Plasmon band diagram along an excursion within the first Brillion zone of the superlattice for $B$=0 (solid curves) and $B$=8 T (dashed curves). (**c-f**) Projected band diagram calculated for a superlattice of finite period ($N$=20) along the zigzag (y direction in Fig. 1a) (**c,d**) and armchair (x direction in Fig. 1a) (**e,f**) boundaries with $B$=0 (**c,e**) and $B$=8 T (**d,f**). The red curves indicate the edge modes on the upper and bottom boundaries. We numerically calculate a Chern number $C$=1 for the band

below the gap with $B$=8 T. The superlattice hexagon side length is $L$=600 nm (see Fig. 1a), while other ribbon parameters are the same as in Fig. 1b.

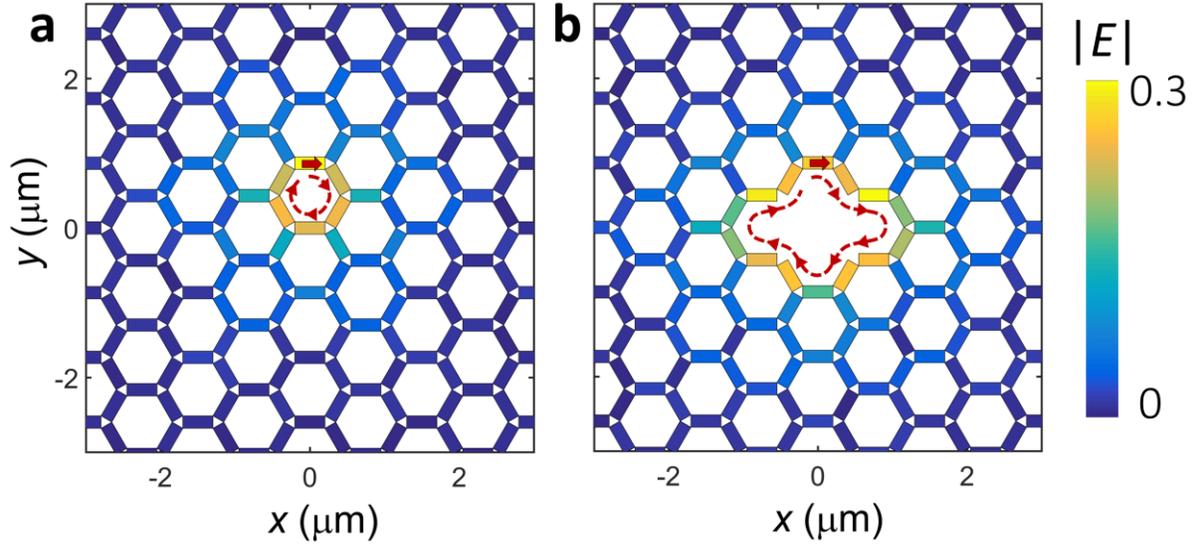

**Figure 3 | Localization of bulk plasmon states.** We show the simulated electric-field distributions induced on the graphene superlattice without (**a**) and with a defect (**b**) upon excitation by a directional point source (red arrow at the center) with a magnetic field of 8 T. The photon energy is at the Dirac point ($E_p$=56.37 meV). The parameters of the superlattice are the same as in Fig. 2.

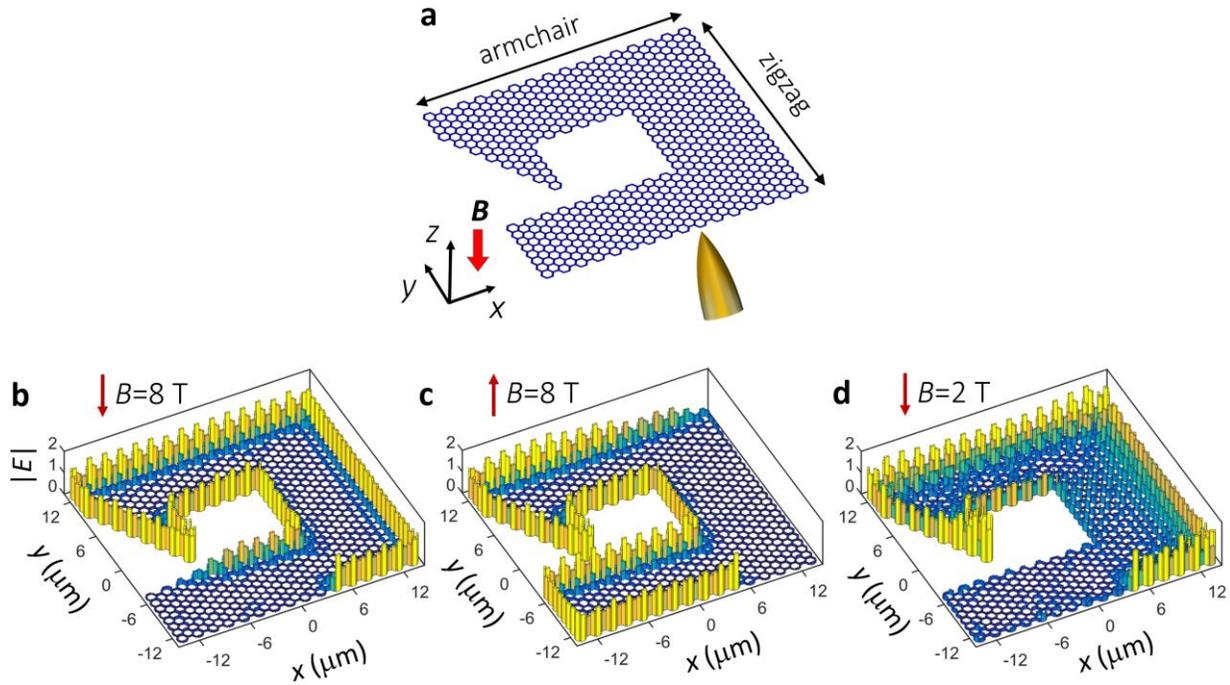

**Figure 4 | Topologically protected edge plasmon.** (**a**) Illustration of a finite graphene superlattice structure with sharp corners, exposed to a normal static magnetic field. A metal tip (lower side) indicates a possible way of exciting the edge mode of this structure. (**b-d**) Optical electric-field distributions in the structure of **a** for a downward magnetic field of either 8 T (**b**) or 2 T (**d**), and for an upward magnetic field of 8 T (**c**). The field distributions are calculated by numerical iteration using the scattering coefficients shown in Fig. 2a. The parameters of the superlattice are the same as in Fig. 2.

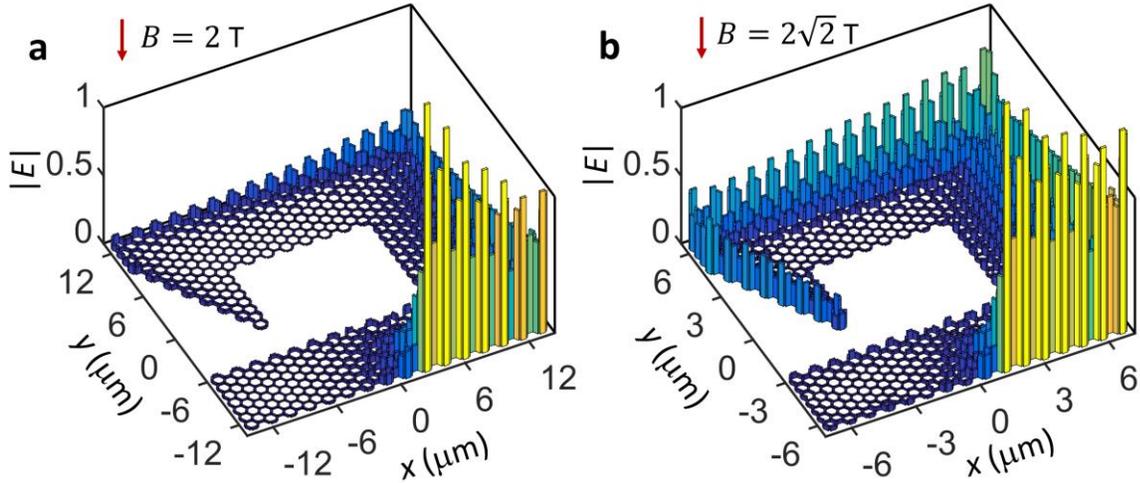

**Figure 5 | The effect of inelastic optical losses.** (**a**) Optical electric-field distributions in the graphene superlattice in Fig. 4 incorporating loss, where the magnetic field is 2 T and the photon energy is $E_0$. (**b**) Edge mode propagating in a superlattice with length size decreased by a factor of 2, assuming a magnetic field of $2\sqrt{2}$ T, for a photon energy of $\sqrt{2}\,E_0$. The graphene mobility is 100,000 cm$^2$/(V·s) in both **a** and **b**.

# Supplementary information

## I. Magneto-plasmon modes on graphene nanoribbon.

A layer of graphene supports bulk plasmons (Fig. S1a) and more localized edge plasmons at the boundary (Fig. S1b). The dispersion relations of bulk plasmons $\omega_B(k)$ (determined by $k = 2i\omega\varepsilon_0/\sigma_0(\omega)$ ) and edge plasmons are correlated: $\omega_E(k) = (2/3)^{1/2}\omega_B(k)$ [1], where the prefactor implies the localization of the edge plasmons. With magnetic field, the dispersion of the bulk plasmons is shifted to $\omega_B'^2 = \omega_B^2 - \omega_c^2$, where $\omega_c$ is the cyclotron frequency defined in the main text, while the edge plasmons become $\omega_{E,\pm}' = \sqrt{2}\left[\left(3\omega_B^2 + \omega_c^2\right) \pm \omega_c\right]/3$ [2-5] , where the signs before $\omega_c$ correspond to the direction of the magnetic field normal to the carbon plane. Because of the symmetry of the system, for the edge magneto-plasmon, reversing the direction of the magnetic field is identical to reversing the direction of the wave vector, $\omega_{E,\pm}'(q) = \omega_E'(\pm q)$. Therefore, the edge magneto-plasmon is nonreciprocal: $\omega_E'(+q) \neq \omega_E'(-q)$, in contrast to the bulk magneto-plasmon, as shown in Fig. S1.

The two lowest-order modes of the nanoribbon are bonding and antibonding combinations of the two edge charge pileups on opposite boundaries of the nanoribbon (Fig. 1b). For a graphene ribbon under a magnetic field, the edge modes on the two boundaries propagating toward the same direction have different wave vectors. The combinations of these two edge modes are analogous to the asymmetric supermodes of two nonidentical coupled waveguides, leading to the results shown in Fig. 1b.

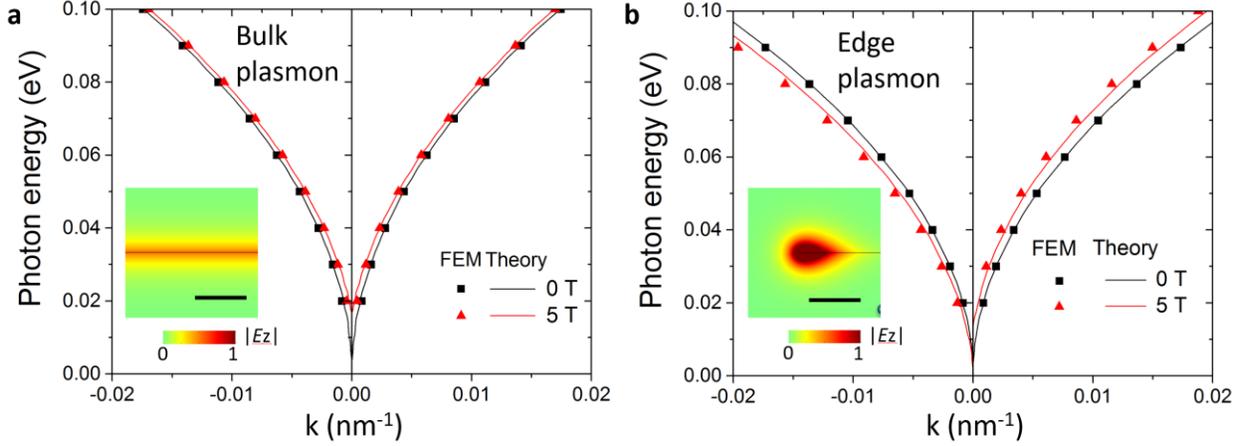

**Figure S1.** Dispersion relations for bulk (**a**) and edge (**b**) plasmons on freestanding graphene. Simulation and theoretical results are shown for zero magnetic field (red) and for a magnetic field of 5 T (black). Insect: distributions of the electric-field component along the propagation direction. Scale bars: 200 nm.

## II. Method for calculation of band diagrams

In general when plasmon waves are simultaneously launched at all three branches of a junction, the amplitude of input and output fields at the three branch terminals are connected by scattering equations as $\left[ A_1^{out} A_2^{out} A_3^{out} \right]^{T} = S \left[ A_1^{in} A_2^{in} A_3^{in} \right]^{T}$ (*i=in, out*), where $S$ is a scattering matrix having three independent variables: $S_{11}=S_{22}=S_{33}$, $S_{12}=S_{23}=S_{31}$ and $S_{13}=S_{21}=S_{32}$, considering the symmetry of the structure. Due to time-reversal symmetry breaking, $S$ is asymmetric for $B \neq 0$ T. Incidentally, out-coupling at the junction to free space is very inefficient due to impedance mismatch (e.g., out-coupling losses are found below 0.01% for a photon energy $E_0$). Neglecting out-coupling the scattering matrix should be unitary, and therefore, energy conservation leads to the condition $|S_{11}|^2 + |S_{12}|^2 + |S_{13}|^2 = 1$.

Using the simulation results of $S_{11}$, $S_{22}$ and $S_{13}$ shown in Fig. 2a, we can calculate the band diagrams of the graphene superlattice. As shown in Fig. S2a, the unit cell of the superlattice (blue dash region) is formed by two nanoribbon junctions. The superlattice is constructed by repetitive translation of the unit cell along the principal axes $\boldsymbol{a_1}$ and $\boldsymbol{a_2}$ shown in Fig. 1a. We label the five pieces of nanoribbons within the unit cell as $i$=1~5. Each of them supports two counter-propagating waves, with field amplitudes of $A_i{}^j$, where $j$=1,2 denotes the directions of the waves corresponding to the red and black arrows in the figure, respectively. The amplitude of these waves are connected by the scattering matrix of the single junction discussed above as $[A_1{}^2\ A_2{}^2\ A_3{}^2]$=$S[A_1{}^1\ A_2{}^1\ A_3{}^1]$ and $[A_3{}^1\ A_4{}^1\ A_5{}^1]$=$\mathbf{S}[A_3{}^2\ A_4{}^2\ A_5{}^2]$. Now, Bloch's theorem on the boundaries leads to $[A_4{}^1\ A_4{}^2]$=$exp(\text{-}i\boldsymbol{K}\cdot\boldsymbol{a_2})[\ A_1{}^1\ A_1{}^2]$ and $[A_5{}^1\ A_5{}^2]$=$exp(\text{-}i\boldsymbol{K}\cdot\boldsymbol{a_1})[\ A_2{}^1\ A_2{}^2]$, where $\boldsymbol{K}$=$k_x\boldsymbol{x}$+$k_y\boldsymbol{y}$ is the Bloch wave vector. The band diagram of the superlattice in Fig. 2b can be obtained from the condition of vanishing determinant of the above linear equations, while the wave functions $|\phi(\boldsymbol{K})\rangle$ are given by the corresponding eigenvectors. We calculate the Chern numbers C of the studied bands by integrating the Berry connection $\langle \phi | \nabla_K | \phi \rangle$ along the boundary of the first Brillouin zone [6].

Using a similar method, we also calculate the projected band diagram for a superlattice of finite width, as shown in Fig. S2b and c. Here, the unit cell is indicated by the two parallel dashed lines, containing a larger number of nanoribbons. Again, each nanoribbon supports two counter-propagating waves. The scattering between these waves is determined by similar relations as above. Here, apart from the coefficients $S_{11}$, $S_{22}$ and $S_{13}$ of Fig. 2a, we also need to obtain the scattering coefficients for V-shape junctions appearing at the boundary of the superlattice. We use the same simulation method as for the three-lobbed junction. The resulting linear set of equations, combined with Bloch's theorem along the direction of periodicity, generate the projected band diagram shown in Fig. 2c-f.

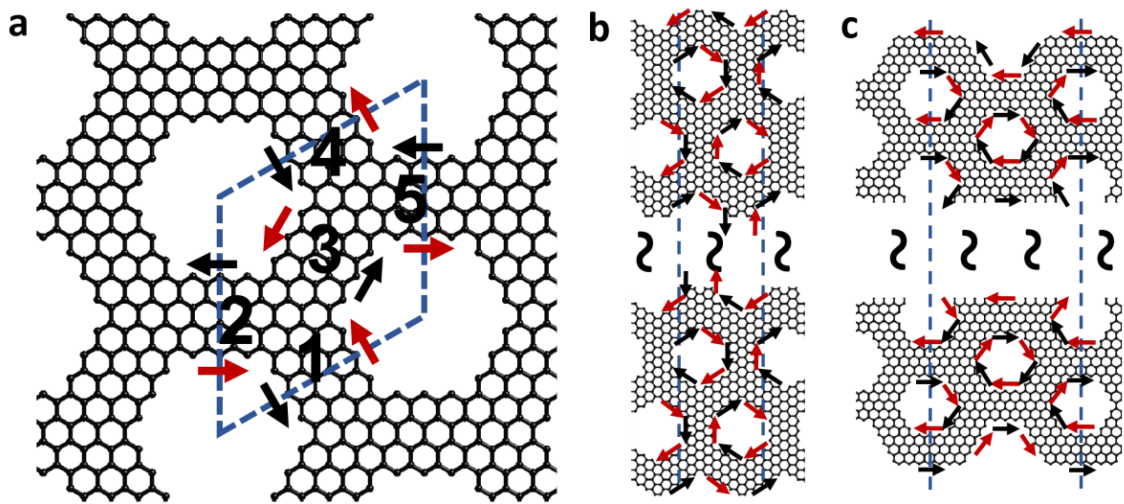

**Figure S2.** Illustrations of an infinite superlattice (**a**) and a superlattice of finite width (**b,c**).

## III. Simulation method for field distribution in a superlattice

We focus on a superlattice of specific shape constructed out of $N$ nanoribbons. We use a $2N$ element vector $A_i(t)$ ($i$=1~$2N$) to describe the distribution of field amplitude on the superlattice at time $t$, with $A_{2n}$ and $A_{2n-1}$ representing the amplitude of two counter-propagating waves at the $n$th nanoribbon. Therefore, $|A_{2n}(t)+A_{2n-1}(t)|$ gives the field amplitude at the ribbons.

For the simulations, we introduce a point source at the initial time and calculate the field distribution in sequential time steps upon iteration. The point source is introduced at the $n$[th] nanoribbon through a prescribed pair of values $A_{2n}(0)$ and $A_{2n-1}(0)$, by setting one of them to 1 (unidirectional point source for Fig. 3) or both of them to 1 (omnidirectional point source for Fig. 4 and 5), with all other elements of $A(0)$ set to 0. Then, the field distribution of the waves scattered by the junctions of the superlattice at a given time can be calculate through $A(1)=TA(0)$, where $T$ is a $2N\times2N$ matrix constructed from the $S_{11}$, $S_{22}$ and $S_{13}$ coefficients (for the bulk), as well as the scattering coefficients for V-shape junctions, which describe scatterings between waves in the superlattice and are determined by the connecting relations of the nanoribbons (see above). Subsequent field distributions can be obtained by iteration $A(t+1)=T[A(t)+A(0)]$, where the initial distribution $A(0)$ is added at every step, so that the point source produces a persistent excitation.

## IV. Electrostatic scaling of graphene plasmons

The general electrostatic potential associated with a plasmon is $\phi(r,\omega)=\left(4\pi\varepsilon_0\bar{\varepsilon}\right)^{-1}\int d^2r'|r'-r|^{-1}\rho(r',\omega)$, where the $\rho(r',\omega)$ is the charge distribution on the graphene, which is related to the surface current through the continuity equation, $\nabla\cdot j(r',\omega)=-i\omega\rho(r',\omega)$, while $\bar{\varepsilon}=(\varepsilon_s+\varepsilon_m)/2$ is the average permittivity of the environment (substrate $\varepsilon_s$ and superstrate $\varepsilon_m$) and a substrate. The current is $j(r',\omega)=\sigma\cdot E(r',\omega)=\sigma\cdot\nabla\phi(r',\omega)$. Therefore, using a dimensionless coordinate vector $\theta=R/W$, the electrostatic potential can be expressed as

$$\phi=\frac{1}{4\pi\varepsilon_0\bar{\varepsilon}}\left[\eta_1\int\frac{d^2\theta'}{|\theta'-\theta|}\left(\partial_x\phi+\partial_y\phi\right)+\eta_2\int\frac{d^2\theta'}{|\theta'-\theta|}\left(\partial_y\phi-\partial_x\phi\right)\right],$$

where $\eta_1=i\sigma_{xx}/\omega W$ and $h_2=iS_{xy}/\omega W$ are functions of the normalized frequencies $\omega/\omega_0$ and $\omega_c/\omega_0$, where $W_0=\sqrt{e^2 E_F/\left(W\,\varepsilon_0\bar{\varepsilon}h^3\right)}$.

The above analysis show that the plasmon fields in the graphene superlattices under study only depend on the geometrical parameters $W$ and $L$, and on the normalized frequencies $\omega/\omega_0$ and $\omega_c/\omega_0$. We provide a numerical corroboration of this electrostatic scaling law in Fig. S3. These simulations demonstrate that sub-tesla magnetic fields are sufficient to generate topologically protected plasmon states. By carrying out FEM simulations similar to those used to obtain Fig. 2a, in which we calculate the scattering coefficients of a single nanoribbon junction (with *L*=3*W*, similar to Figs. 2-5), but with different combinations of parameters as a function of $\omega/\omega_0$, with W adjusted to ensure that $\omega_c/\omega_0$=0.31 is same as in Fig. 4d. The simulation results show that the scaling law is perfectly fulfilled for different ribbon widths, doping levels, substrate environments and strengths of the magnetic field. These results also indicate that a superlattice of larger size (*W*=806 nm with *L*=3*W*) operating under a magnetic field of only 0.8 T and a plasmon energy of 14 meV exactly resembles the results of Fig. 4d. Therefore, the required magnetic field can be further decreased into sub-tesla regime when the lower-energy plasmons are considered.

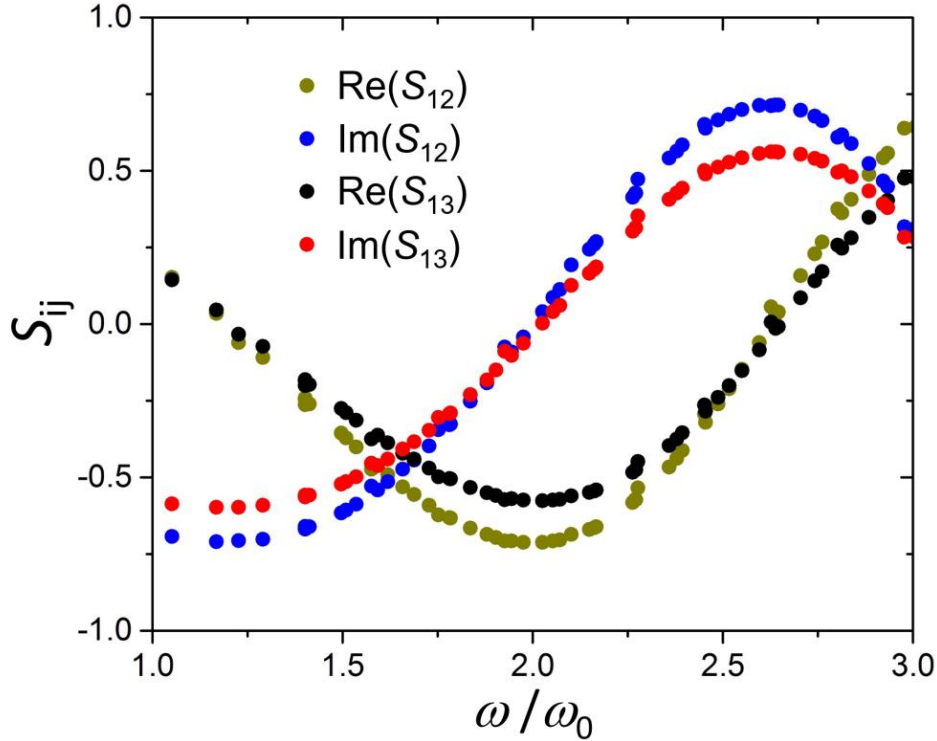

**Figure S3.** Electrostatic scaling of the graphene superlattice. We show the scattering coefficients of graphene nanoribbon junctions (with 2*L*=3*W* as in the superlattice of Figs. 2 and 3) for all combinations of the parameters $B$=0.8, 2 T, $E_F$ =0.2, 0.3 eV, and $\varepsilon_2$=1, 2.1, as a function of $\omega/\omega_0$. The size is adjusted to ensure that $\omega_c/\omega_0$ is the same as in the right panel of Fig. 4d for all calculations.